# Stern-Gerlach deflection of cryogenically cold polyatomic molecules in superfluid nanodroplets

*Benjamin S. Kamerin, Thomas H. Villers, John W. Niman, Jiahao Liang, Angel I. Pena Dominguez, Vitaly V. Kresin*[*]

Department of Physics and Astronomy, University of Southern California; Los Angeles, 90089-0484, USA

**Abstract:** Beam deflection is capable of providing valuable information about the magnetic moments of molecules and clusters as well as the relaxation dynamics of their spins. However, observations have been hampered by magnetic couplings to excited vibrational and rotational states of polyatomic systems, which are challenging to control, characterize, and systematize. In this work, we carried out deflection measurements on superfluid helium nanodroplets doped with high-spin $FeCl_2$ and $CoCl_2$ molecules and their complexes. This enabled quantitative determination of the magnetic moments of molecules and clusters at extremely low, and fully defined, temperature of all of their degrees of freedom. The spin magnetic moments become thermalized and oriented along the applied field. Dimers and trimers are found to be antiferromagnetically ordered. The issue of rates and mechanisms of molecular spin relaxation within the cryogenic helium matrix is highlighted.

[*] Email *kresin@usc.edu*

The Stern-Gerlach experiment,[1,2] performed just over 100 years ago, not only revolutionized physics but also pioneered the technique of beam deflection by an inhomogeneous magnetic field.[3] This powerful method of probing the quantum properties of individual atoms isolated from external perturbations has been extensively applied to the study of their electronic and nuclear magnetic moments.[4-6] A natural subsequent step was to extend it to beams of diatomic [7,8] and polyatomic[9-12] molecules, and metal[13,14] and molecular[12,15,16] nanoclusters.

However, for systems with many different degrees of freedom magnetic deflection data become substantially more complex to interpret. The electrons' orbital and spin angular momenta couple to the molecular framework in different ways, and these interactions and their anisotropies depend on the symmetries and populations of the rotational and vibrational states. Even for the same molecule or cluster, its net magnetic moment may behave as if fully locked to the framework, as thermally fluctuating, or as diffusing in an intermediate fashion, depending on the relevant temperature.[17] This variability can translate, for example, into quite different shapes of the measured beam profiles (e.g., uniformly deflected, symmetrically broadened, segmented, undeflected) of the same molecule or cluster when studied under different conditions[17] or in different setups.[9,18]

The problem of internal vibrational and rotational temperatures, which does not arise in atomic Stern-Gerlach measurements, poses a crucial challenge in experiments on clusters and molecules in beams. Furthermore, rotational and vibrational excitation suppresses magnetization, making deflections difficult to resolve. Supersonic expansions,[6] buffer gas cooling,[19] and cryogenic nozzles[20] cannot easily and universally achieve subkelvin temperatures, especially for polyatomic molecules and clusters. Also problematic is the fact that in beams the translational, rotational, and vibrational degrees of freedom commonly cool to a highly dissimilar degree and even to a non-thermal distribution. [17]

In this work, we present a method to obtain magnetic deflection data on fully thermalized, extremely cold molecules and molecular clusters. Uncertainties related to the particles' internal state are thereby eliminated. The resulting Stern-Gerlach profiles can be employed for a direct, well-defined determination of magnetic moments. This is achieved by first embedding the molecules of interest within superfluid helium nanodroplets and then deflecting the doped nanodroplet beam. Inside these "personal flying cryostats" the molecules are cooled to their lowest vibrational and low rotational states.

It is important to emphasize that probing cold molecules by nanodroplet embedding is applicable to a broad range of systems, from atoms to polyatomic molecules to complexes assembled and cooled inside the nanodroplets by sequential pickup of either identical or distinct species.[21,22] In a recent publication[23] we outlined a measurement on a diatomic molecule (metastable $Na_2$). The present work is the first application to high-spin polyatomic molecules and their clusters. The orientation, relaxation, and interaction of spin magnetic moments in such a cryogenic environment are also of significant interest in the context of research on molecular magnets and molecular qubits.[24]

Our apparatus produces a beam of $He_N$ nanodroplets by expansion of high-pressure helium gas through a cold nozzle, see the supplementary material for details. Upon leaving the nozzle the nanodroplets undergo prompt evaporative cooling, attaining an internal temperature of 0.37 K.[21] Then they pass through dilute molecular vapor inside a heated pickup cell where they readily embed molecules of interest. As is the case for impurities other than the alkalis,[25] the dopant molecule moves to the center of the droplet and becomes fully thermalized by dissipating its internal and translational energy into the liquid. Thanks to the high thermal conductivity of liquid helium,[26] the released energy is transferred to the nanodroplet surface, and the internal temperature is restored by prompt evaporation of additional helium atoms. Furthermore, the $^4$He matrix is completely nonmagnetic, and the nanodroplets are superfluid, allowing the dopants to easily rotate and reorient themselves in an external field.[21,22]

As dopants, we employ $FeCl_2$ and $CoCl_2$. In the bulk, above their Néel temperatures of 24-25 K,[27] these are well known as paramagnetic salts [28] whose magnetic response derives from the metal ions' unfilled *d* shells. Their susceptibility obeys Curie's law, but the orbital angular momentum is quenched by the crystal field splitting and therefore the magnetic moment is defined by the electron spin. By contrast, there seem to have been no studies of the magnetic properties of individual constituent molecules. This makes them an attractive subject for a Stern-Gerlach measurement, as do their large anticipated magnetic moments. The latter are advantageous in the current setup because the embedded molecule must experience a sufficiently strong force from the applied field gradient in order to measurably deflect its heavy host nanodroplet.

After collimation to the shape of a narrow ribbon, the doped nanodroplet beam passes between the poles of the deflection magnet. As described in Ref. 29, the deflector (magnetic field 1.1 T, field gradient 330 T/m) is powered by permanent rare-earth magnets, which enables a more compact and lightweight design. This is followed by a free flight region where the small (~1 mrad) angular deflection of the beam is translated into a measurable lateral displacement. Finally, the nanodroplets enter the electron-impact ionizer of a quadrupole mass spectrometer through another collimating slit. Here the ionized dopants and their fragments are ejected from their nanodroplets,[30] mass filtered, and detected by an ion counter. By mounting the detector chamber on a precision translation stage, we map out the transverse intensity profile of the doped nanodroplet beam.

The deflection induced by the magnetic field is measured by comparing the beam's "magnet-in" and "magnet-out" profiles. The amount of deflection is defined by the dopant's magnetic moment, the nanodroplet mass, and the beam velocity. The nanodroplet size distribution, known to be log-normal,[21,22] is determined during the same experimental run by separately doping them with a polar molecule (here, CsI) and performing an electric deflection measurement.[31] The beam velocity distribution is extracted from a time-of-flight measurement.[32] Conveniently, this distribution is very narrow because it is formed in a supersonic expansion.

**Deflection of nanodroplets containing individual molecules**

The deflection profiles of nanodroplets doped with $FeCl_2$ and $CoCl_2$ are shown in Fig. 1. A clearly resolved one-sided (paramagnetic) deflection of the beam is observed. With the nanodroplet size distribution, velocity distribution, and the magnetic field gradient all known, the dopants' magnetic moments are obtained by a simulation of the deflection process, as described in the supplementary material. Importantly, the Brillouin susceptibility function[28] certifies that at the nanodroplet temperature of 0.37 K the immersed magnetic moment is fully oriented by the applied magnetic field.

From these data, the magnetic moments are determined to be $5.5 \pm 0.5$ $\mu_B$ (Bohr magnetons) for $FeCl_2$ and $4.7 \pm 0.5$ $\mu_B$ for $CoCl_2$. Remarkably, these values are essentially identical to the crystalline bulk environment: 5.4 $\mu_B$ and 4.8 $\mu_B$, respectively.[28]

The results immediately lead to several nontrivial implications:

(i) Stern-Gerlach deflection of beams of superfluid helium nanodroplets with entrapped magnetic impurities is viable and measurable. This opens the door to quantitative determination of magnetic moments of a variety of molecules and clusters in a state of extremely low and fully defined temperature of all of their degrees of freedom.

(ii) The electronic configurations and ground state terms for the atomic ions $Fe^{2+}$ and $Co^{2+}$ are $3d^6$ ($^5D_4$) and $3d^7$ ($^4F_{9/2}$), respectively. Nevertheless, as mentioned above, the magnetic susceptibility of bulk salts containing iron-group ions reveals that they behave as if only their spin moments contribute to magnetization (i.e., as if $J \approx S$).[28] This is explained by the phenomenon of orbital angular momentum quenching, that is, the complete splitting of the degeneracy of the orbital $L$ multiplet by the crystal field. Since the molecular magnetic moments measured here are practically identical to the bulk values, it follows, interestingly, that already in the individual triatomic molecules the quenching is as effective as in the full crystalline lattice.

This implies, in particular, that the submerged molecules cannot be strictly linear.[34-36] Indeed, in a symmetric linear molecule the central ion would retain the degeneracy of its states with opposite projections of the orbital angular momentum.[37] The expectation value of $\hat{\vec{L}}$ and the orbital magnetic moment would then remain non-zero,[38] resulting in a markedly different value of total magnetic moment $\mu$.

(iii) The consistency of all the experimental data reported here demonstrates that the spin magnetic moments are fully thermalized to the nanodroplet temperature. It follows that the spin relaxation time is much shorter than the ~ 300 μs beam flight time through the deflector.

That this should be the case is far from obvious. While, as noted above, phonon and ripplon nanodroplet excitations[22] efficiently remove the energy of the dopants' vibrations and rotations, $^4He$ is completely nonmagnetic and therefore cannot directly affect the electron spin. Indeed, for alkali-metal atoms the longitudinal spin relaxation time in both solid[39] and superfluid[40] helium is very long: 1-2 seconds. Similarly, for alkali atoms residing on the surface of helium nanodroplets their spin relaxation was found to proceed at a longer than millisecond time scale.[41,42]

Therefore it is apparent that there exists a pronounced difference between atomic and molecular spin relaxation. In fact, metastable high-spin alkali dimers and trimers on nanodroplet surfaces also demonstrated short relaxation rates, below a microsecond.[41,42] What pathway, then, may enable prompt thermalization in the present case of ionic molecules residing in the nanodroplet interior?

In solid-state electron spin resonance relaxation predominantly proceeds via the spin-phonon interaction.[43] However, for our individual triatomic dopants this is impossible: the vibrational normal modes are frozen out, and their frequencies in $FeCl_2$ and $CoCl_2$, 10-60 meV,[34,35] greatly exceed the Zeeman energies (0.3-0.4 meV), i.e., there are no soft phonon modes into which the spin-flip energies could be dissipated. Consequently, in the present case it is the spin-rotation interaction[44-46] that can be presumed to facilitate thermalization. The rotational constant of $FeCl_2$, for example, is 0.048 $cm^{-1}$ = 6 μeV,[36] hence energy transfer to molecular rotations, with subsequent dissipation into the helium bath, is quite feasible.

A related question is the strength of the spin-rotation coupling. This is a complex problem that does not appear to have been analyzed theoretically for the systems in question. As discussed in Ref. 44, for an order-of-magnitude estimate two coupling mechanisms should be taken into account. One is the interaction of the metal cation's magnetic dipole moment with the magnetic field of the circular current generated by the two rotating chlorine anions. Using the molecule's dimensions and an estimate of its rotational velocity at the nanodroplet temperature, magnetostatics yields a coupling energy of $\Delta_1 \sim 10^{-9}$ eV. The other coupling arises from Van Vleck's perturbation theory treatment of the combined action of the electron spin-orbit interaction and of the interaction of electron orbital angular momentum with molecular rotations;[47,48] its order of magnitude is $\Delta_2 \sim \zeta B/\varepsilon$. Here $\zeta$ is the spin-orbit coupling strength, $B$ the rotational constant, and $\varepsilon$ the energy interval between electronic states of different angular momenta. Taking $\zeta \sim \varepsilon \sim 10$ meV[27,34] yields $\Delta_2 \sim 10^{-6}$–$10^{-5}$ eV. Clearly, these values are only schematic and insight from a quantitative theoretical study will be of value. It is interesting, though, that if the above magnitude of $\Delta_1$ is used as an estimate of the energy broadening of the molecule's spin in the external magnetic field, the corresponding lifetime is $\Delta t \sim \hbar/\Delta_1 \sim$ 1-10 μs, which is a plausible value for the spin relaxation timescale in the present experiment.

### Deflection of nanodroplets containing molecular assemblies

By gradually increasing the vapor density in the pickup cell, we can produce nanodroplets doped with multiple molecules, and by utilizing two pickup cells in a row we can produce mixed complexes. Such step-by-step exploration of the magnetic interaction between adjacent molecules is informative because in the bulk at low temperatures the corresponding materials have a famously intriguing magnetic behavior: ferromagnetic intraplane coupling, antiferromagnetic interplane coupling, and a "metamagnetic" field-dependent behavior of the magnetization.[49-51]

Figure 2(A) reveals that the dimer $(FeCl_2)_2$ exhibits no discernible deflection. The calculation in Ref. 52 identified two low-energy conformations of this dimer, one antiferromagnetic ($\mu = 0$) and the other ferromagnetic ($\mu = 8\mu_B$). The former was predicted to be energetically favorable, by 0.04 eV. Now the deflection measurement is able to unequivocally confirm this assignment. Indeed, the presence of even a relatively small population of the ferromagnetic conformer would have been made apparent by the appearance of a strongly deflected fraction of the beam. (The formation of other metastable isomers is not likely to proceed within a nanodroplet because it is energetically unfavorable and has been observed only for highly polar molecules.[53,54])

Figure S4 in the supplementary material shows that the magnetic moment of the $(CoCl_2)_2$ dimer is likewise undetectable, while that of the mixed cluster $(FeCl_2)\cdot(CoCl_2)$, $0.7 \pm 0.5\mu_B$, is small and fully consistent with an opposing orientation of the two dipoles.

Proceeding to embedded trimers, the deflection profiles of $(FeCl_2)_3$ and $(CoCl_2)_3$ are shown in Fig. 2(B) and in the supplementary material in Figs. S5, S6. The magnitudes of their magnetic dipoles ($5.7 \pm 0.5\mu_B$ and $4.6 \pm 0.5\mu_B$, respectively) are the same as the monomers'. Thus the deflection measurement reveals that the molecules continue to assemble antiferromagnetically.

Having been shown to be practical and quantitative, the nanodroplet deflection technique can be applied to a variety of atomic, molecular, and cluster systems. For example, it will be possible to explore the magnetic ordering of larger molecular assemblies and unusual complexes formed within the nanodroplet environment, and to investigate whether adding electron donors or acceptors to such complexes can transform them into strongly deflecting ferromagnetic configurations.[52] Understanding the dependence of spin coherence and thermalization time scales on the size and composition of the embedded system is another question of fundamental interest. Similarly, extending cluster deflection experiments to the well-thermalized nanodroplet setting also promises to be informative, as it will eliminate uncertainties regarding thermal dynamics and elucidate the interplay of magnetism with structural features and with geometric and electronic "magic numbers".[13,55,56] The challenge with such an extension lies in being able to identify a specific cluster's signal in an increasingly congested daughter fragment ion mass spectrum. It may be possible to accomplish this with the aid of high-resolution electron-attachment mass spectrometry in combination with isotopic fitting.[57,58]

*Supplementary material* provides details on the experimental arrangement and methods and on fitting the deflection profiles and error estimation. It also includes plots of droplet beam velocities, sample mass spectra, and the deflection profiles of $(CoCl_2)_2$, $(FeCl_2)\cdot(CoCl_2)$, $(CoCl_2)_3$, and of $(FeCl_2)_3$ acquired at an alternative fragment ion mass peak.

*Acknowledgment.* This work was supported by the U.S. National Science Foundation (grant No.CHE-2153255).

*Conflict of interest.* The authors have no conflicts to disclose.

*Author contributions*. **B. S. Kamerin**: Conceptualization (equal); Data curation (lead); Formal analysis (lead); Investigation (lead); Methodology (equal); Software (equal); Writing – review & editing (lead). **T. H. Villers**: Investigation (equal); Writing – review & editing (equal). **J. W. Niman**: Conceptualization (equal); Investigation (equal); Methodology (equal); Software (lead); Writing – review & editing (equal). **J. Liang**: Investigation (supporting); **A. I. Pena Dominguez**: Investigation (supporting); Writing – review & editing (equal); **V. V. Kresin**: Conceptualization (lead); Formal analysis (equal); Funding acquisition (lead); Methodology (equal); Writing – original draft preparation (lead).

*Data availability.* The data that support the findings of this study are available from the corresponding author upon reasonable request.

**Fig. 1. Deflection profiles for superfluid helium nanodroplets doped with (A) $FeCl_2$ and (B) $CoCl_2$.** Open circles and the centered (black) lines are the undeflected beam profile data and their smoothing fit, respectively. Solid circles are the deflected beam profile data, and the colored lines are simulations of the deflection process. From the latter, the molecule's magnetic moment is determined. The profiles were mapped out by setting the mass spectrometer detector to the $FeCl^+$ and $CoCl^+$ fragment ion peaks, see Refs. 31-33 and the supplementary material for details. The deflected profiles are slightly asymmetrical because smaller nanodroplets from the log-normal size distribution in the beam deflect stronger than larger ones. The average size of nanodroplets emitted by the nozzle in these experiments was $5.5–6.5\times10^3$ He atoms, but the average size of nanodroplets large enough to pick up and retain a molecule was approximately 60% larger.

**Fig. 2. Deflection profiles for superfluid helium nanodroplets doped with (A) the dimer $(FeCl_2)_2$ and (B) the trimer $(FeCl_2)_3$.** The symbols are the same as in Fig. 1. The profiles were mapped out by setting the mass spectrometer detector to the $Fe_2Cl_3^+$ and $FeCl_3^+$ fragment ion peaks of the dimer and the trimer, respectively. (The assignment of these ions to their parent clusters has been verified by mass spectrometry and cross-checked with the deflection of other ions, as described in the supplementary material.) The average size of nanodroplets emitted by the nozzle in these experiments also was $\approx 6\times10^3$ He atoms, while the average size of nanodroplets large enough to pick up and retain a trimer was approximately three times larger.

*SUPPLEMENTARY MATERIAL FOR*

# Stern-Gerlach deflection of cryogenically cold polyatomic molecules in superfluid nanodroplets

*Benjamin S. Kamerin, Thomas H. Villers, John W. Niman, Jiahao Liang, Angel I. Pena Dominguez, Vitaly V. Kresin*[*]

Department of Physics and Astronomy, University of Southern California; Los Angeles, 90089-0484, USA

## S1. EXPERIMENTAL ARRANGEMENT

A diagram of the experimental apparatus is shown in Fig. S1. In the nanodroplet source, helium of research grade purity expands through a cryocooled 5 μm nozzle. In the experiments described in this work, the helium stagnation pressure in the source was maintained at 30 bar and the nozzle temperature was 14 K.

The beam passes through a 0.2 mm diameter skimmer followed by a rotating wheel chopper and two consecutive heated pickup cells. Each cell provides a 2.75 cm long path through the vapor. The first one is filled with either $FeCl_2$ or $CoCl_2$ powder, and the second one with CsCl powder used for nanodroplet size calibration. All cells and their contents are baked and outgassed prior to the measurements.

The doped beam then enters the 15 cm long deflection magnet through a 0.37 mm wide × 2.22 mm high slit. This is followed by a 1.25 m free flight region, after which the beam passes through a 1.4 mm wide slit into the electron impact ionizer of a quadrupole mass spectrometer (Ardara Technologies) which is mounted on a horizontal translation stage driven by a stepper motor. The detector is scanned through a randomized sequence of positions, mapping out the spatial profile of the incoming beam.

The ionization energy is set to 70 eV. The mass spectrometer is equipped with a pulse counting channeltron, and the dopants' fragment ions are counted in synchronization with a beam chopper. This synchronization of the pulse counters allows reliable detection even when beam intensity becomes highly attenuated by the collimation, down to ion signals as low as several counts per second. The ion fragment mass spectra are discussed in Section S4.

## S2. DEFLECTION MAGNITUDE AND VELOCITY MEASUREMENT

If the deflecting field and its gradient point in the $z$ direction, the magnitude of beam deflection is given by[29]

$$d_z = \frac{l_1(\frac{1}{2}l_1 + l_2)}{mv^2}\langle \mu_z \rangle \frac{dB}{dz} \tag{S1}$$

Here $m$ is the mass of the doped nanodroplet, $v$ is the beam velocity, $l_1$ is the full length of the magnet, and $l_2$ is the field-free flight distance between the exit of the magnet and the detector entrance. The field gradient was previously calibrated by a measurement using an atomic beam.[29] The quantity $\langle\mu_z\rangle$ is the projection of the magnetic moment on the field axis. Since in the cryogenic nanodroplet the dopants' magnetic moments become aligned with the external magnetic field, we can set $\langle\mu_z\rangle = \mu$, the magnetic moment in the ground electronic and vibrational state of the molecule.

It is clear from Eq. (S1) that a quantitative deflection experiment requires accurate knowledge of nanodroplet velocities. The fact that the nozzle expansion produces a supersonic beam with a very narrow velocity spread is highly beneficial in this regard. Velocity measurements are performed by using the beam chopper in combination with a multichannel scaler,[32] see Fig. S2.

## S3. MODELING THE DEFLECTION PROFILE: DETERMINATION OF NANODROPLET SIZES AND MAGNETIC MOMENTS

The magnetic moment $\mu$ transported by the nanodroplet beam is deduced by starting with the undeflected beam profile (measured with the magnet retracted and an identical collimator positioned in the beam path), modeling the deflection of each of its slices for an assumed value of $\mu$, and finding the value which provides the best match to the measured deflected ("magnet in") beam profile.

Simulation of the deflection profile. The simulation is described in full detail in Refs. 31-33. It begins by taking the undeflected profile as input (which implicitly accounts for the transverse velocity distribution of the helium nanodroplets as well as for geometrical effects that affect the beam between the source and the detector). For computational convenience and to smooth and extrapolate the data, this profile is fitted to a pseudo-Voigt function (shown as the black lines in Figs. 1, 2, and S4–S6). This zero-field profile curve is then broken into uniform intervals, with the intensity at each position proportional to the number of droplets which will be simulated to deflect from this position.

The simulation uses a Monte Carlo method, which starts by drawing a nanodroplet at random from a log-normal distribution of sizes (described below) and then follows it through the stages of (i) dopant pickup; (ii) deflection by the magnetic field; (iii) spatial filtering by the detector entrance slit; (iv) electron-impact ionization.

Stage (i) accounts for the fact that pickup events occur with a Poisson probability distribution defined by the product of the droplet geometric cross section ($\sigma \propto N^{2/3}$, where $N$ is the number of helium atoms), dopant vapor density, and pickup cell length.[21,22] Furthermore, each such event results in the droplet evaporating atoms and shrinking due to dissipation of the dopants' energy (see below). Stage (ii) was described in Section S2. Stage (iv) takes into consideration the fact that the efficiency of electron-impact ionization and the probability of positive charge subsequently migrating to the dopant are both dependent on the droplet size. The former is also proportional to the cross section of the (shrunken) droplet entering the ionizer, and the latter is proportional to $\exp(-\gamma N^{1/3})$. The parameter $\gamma$ is related to the mean free path of electron hopping in the nanodroplet and has been found[31] to equal ≈0.1.

Nanodroplet evaporation. Boiloff of helium atoms from the surface follows every pickup event in stage (i) and leads to a stepwise shrinkage of the droplet diameter. As described in the main text, in this way evaporative cooling promptly restores the internal nanodroplet temperature. The number of lost atoms is given by the ratio of the energy deposited by the dopant to the energy released per one atomic evaporation (0.5-0.6 meV[21]).

The amount of energy dissipated into the nanodroplet is composed of the molecule's translational, vibrational, and rotational energy at the vapor temperature in the pickup cell, and of the energy released when dopants combine into a complex inside the nanodroplet. The translational kinetic energy lost in the inelastic collision between a molecule and a nanodroplet is the average over all relative collision angles and molecular speeds. An analytical expression for this quantity has been derived from the kinetic theory of gases.[S1] The vibrational energy is equal to the thermal energy of quantum oscillators corresponding to the normal modes of $FeCl_2$ and $CoCl_2$: the symmetric and antisymmetric stretches and the two-fold degenerate bend.[34,35] The rotational constants of the molecules are low[36] and therefore their rotational energies are given by the classical value for a quasilinear rigid rotor. Finally, the complexation energy of monomer addition for all cases was approximated by the dissociation enthalpy of gaseous $FeCl_2$ given in Ref. S2.

Nanodroplet sizes. As mentioned in the main text, the population of nanodroplet sizes produced by the source is determined during each experimental run by doping them with CsI loaded into a separate pickup cell and performing an electric deflection measurement.[31-33] Since the electric dipole moment of this molecule is known, the same simulation as described above is used to fit the parent nanodroplet size distribution to the electric deflection data. (The degree of droplet shrinkage upon capture of the CsI molecule is calculated analogously to the description above, using its own molecular parameters and its pickup cell temperature.) For the conditions used in the present experiments, the typical mean value of the distribution was $\bar{N} \approx 7000$ atoms, and its width was well described by $\Delta N = 0.9\,\bar{N}$.

Of the nanodroplets in this initial distribution, not all are equally likely to pick up a specific number of dopants: the larger the geometric cross section the higher the probability of pickup collisions.[25] Therefore the average size and mass of the nanodroplets carrying 1, 2, or 3 molecules becomes progressively larger that the $\bar{N}$ above. This explains why, for example, the deflection of the $(FeCl_2)_3$ trimer in Fig. 2(B) is smaller than that of the monomer in Fig. 1(A) despite them having the same magnetic moments. The average sizes of the corresponding subpopulations with a given number of dopants is given in the figure captions.

Error estimation. As seen from Eq. (S1), the accuracy of determining a magnetic moment from the deflection profile depends on the quality of profile fit to the data and on uncertainties in the nanodroplet mass distribution, beam velocity, and the magnetic field gradient.

The contributions of the last two parameters are small. As described in Section S2, the beam velocity is measured accurately and has a very small spread, while the magnetic field gradient has been calibrated via a Stern-Gerlach measurement on an atomic beam.[29] The main sources of error lie (i) in the scatter of experimental data points which define the beam profile; and in effects related to nanodroplet sizes: (ii) the determination of the average nanodroplet size in the nozzle beam (see the subsection on “Nanodroplet sizes” above) and (iii) the precise dependence of the efficiency of dopant ionization on droplet size (see the subsection on “Simulation of the deflection profile” above). We found that these two latter effects lead to a ±250 atom uncertainty from each in the average nanodroplet size. Simulations of the deflection process showed that each of the sources (i-iii) contributes an uncertainty of approximately $0.3\mu_B$ to the outcome. Combining these in quadrature, we arrived at the overall estimated experimental uncertainty of $\pm 0.5\mu_B$.

## S4. Dopant ion fragments in the mass spectra

As described above, molecules are picked up by helium nanodroplets via successive collisions in a Poisson process. Therefore it is important to correlate measured beam deflections with specific sizes of the molecular complex. Dopants within nanodroplets are ionized indirectly via charge transfer to $He^+$ resulting from electron bombardment; this transfer is a highly exothermic process which can cause fragmentation.[30] Consequently, when mapping out the deflection profile of a dopant ion peak in the mass spectrum, we need to ensure that it is not a fragment of a larger agglomerate. This is done by gradually increasing the vapor pressure in the pickup cell and monitoring the mass spectrum for the appearance of molecular ions characteristic of progressively larger entities.[32,33] Fig. S3 illustrates thus with a series of representative mass spectra for the $CoCl_2$ dopant.

Having identified the dopant parent ions and their daughter fragment ions, we set the quadrupole mass filter to those mass channels which offer the optimal combination of intensity and separation both from the nearby pure $He_N^+$ peaks and from the fragments of smaller dopant complexes (if present). For example, for Figs. 2(B) and S5 the deflection profiles were collected at the masses of the $FeCl_3^+$ and $CoCl_3^+$ fragments, as they were ascertained to derive from trimer complexes. Fig. S6 confirms that different fragments of the same parent indeed show identical deflections.

## FIGURES

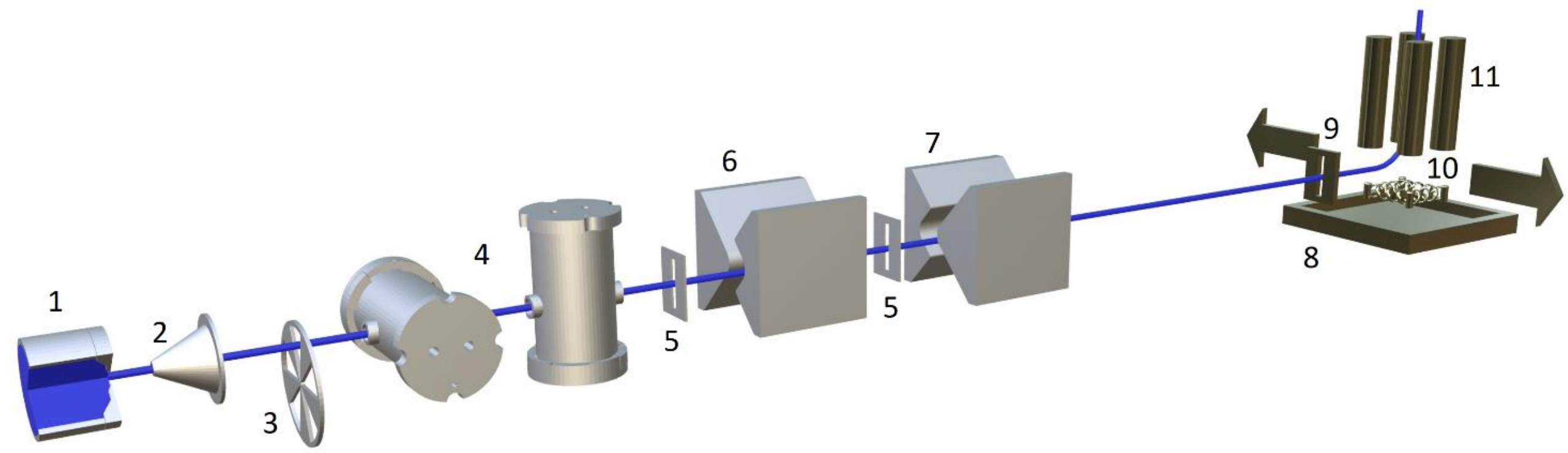

**Fig. S1.** An outline of the experimental apparatus (not to scale). 1: Helium expansion nozzle, 2: Skimmer, 3: beam chopper, 4: pickup cells, 5: electric and magnetic deflector entrance collimators (moved out of the beam path during magnetic and electric deflection measurements, respectively), 6: electric deflector, 7: magnetic deflector, 8: detection chamber on a horizontal translation stage, 9: detector entrance slit, 10: electron-impact ionizer, 11: quadrupole mass spectrometer with a pulse-counting channeltron ion detector.

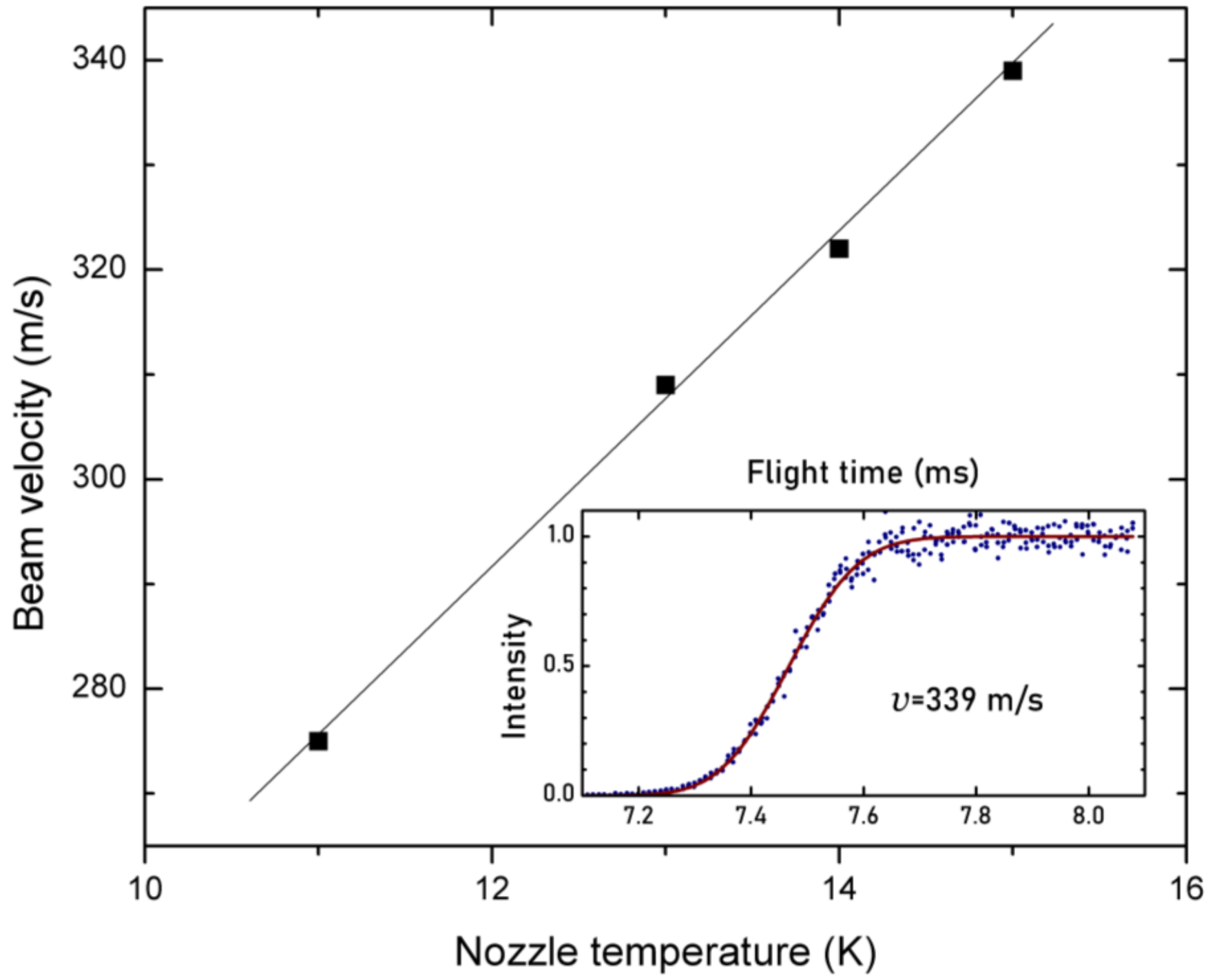

**Fig. S2.** Velocity of the nanodroplet beam for helium source stagnation pressure of 30 bar and nozzle temperatures between 11 and 15 K. The width of the velocity distribution was approximately 2%. These parameters were derived from time-of-flight measurements of beam travel from the chopper to the detector, with an example shown in the inset.

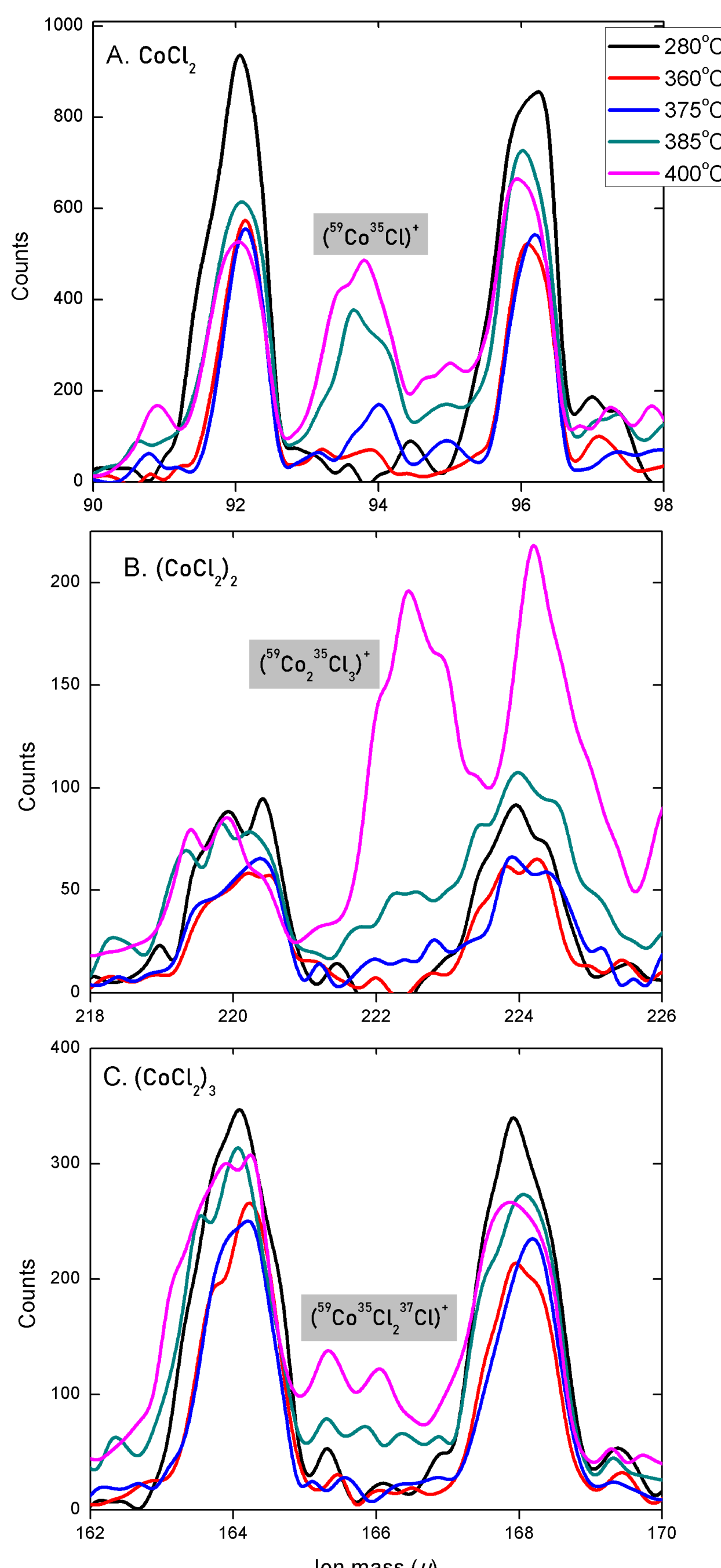

**Fig. S3.** Sample smoothed mass spectra of the fragment ions of $CoCl_2$ complexes formed within helium nanodroplets, identified as deriving from (A) monomers, (B) dimers, (C) trimers. Prominent neighboring $He_N^+$ peaks aid in the mass assignment of the fragments. As the pickup cell temperature increases, some of the isotopologues begin to overlap with the helium peaks. Different line colors correspond to different pickup cell temperatures (see the legend). By adjusting the temperature, monomer signals usable for a deflection measurement can be obtained before the dimers display a significant count rate, and similarly for dimers vs. trimers.

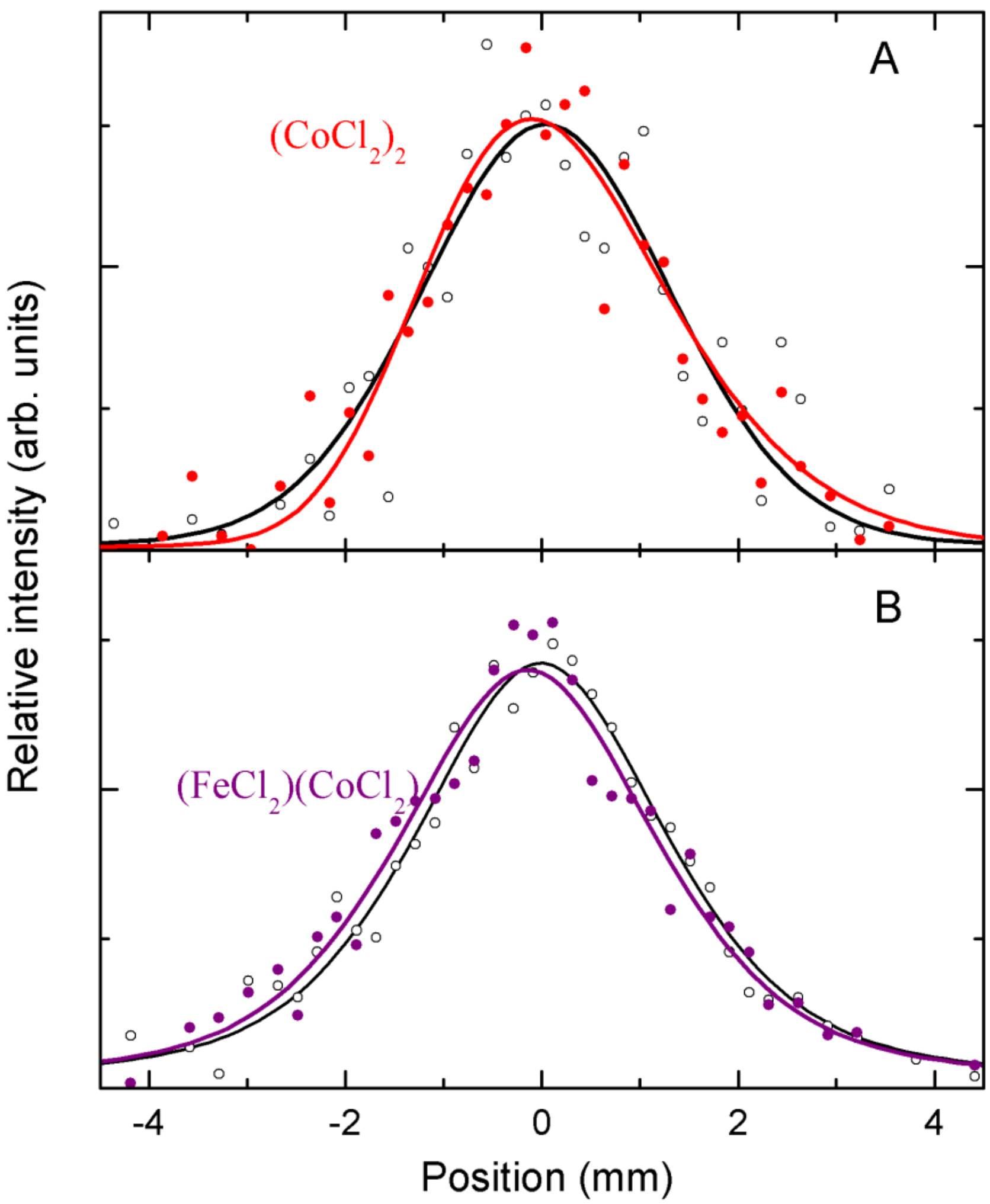

**Fig. S4.** Deflection profiles for nanodroplets doped with (A) the dimer $(CoCl_2)_2$ and (B) the mixed dimer complex $(FeCl_2)\cdot(CoCl_2)$. The symbols are the same as in Fig. 1 in the main text. The profiles were mapped out by setting the mass spectrometer detector to the $Co_2Cl_3^+$ and $FeCoCl_2^+$ fragment ion peaks. The average size of nanodroplets emitted by the nozzle in these experiments was $7\times10^3$ He atoms, while the average size of nanodroplets large enough to pick up and retain a mixed dimer was approximately twice as large. No deflection is resolved for $(CoCl_2)_2$, consistent with a zero magnetic dipole moment. The small deflection of $(FeCl_2)\cdot(CoCl_2)$ yields a fit of $0.7\pm0.5\mu_B$, which matches the value expected for antialignment of the two molecules' spin dipoles.

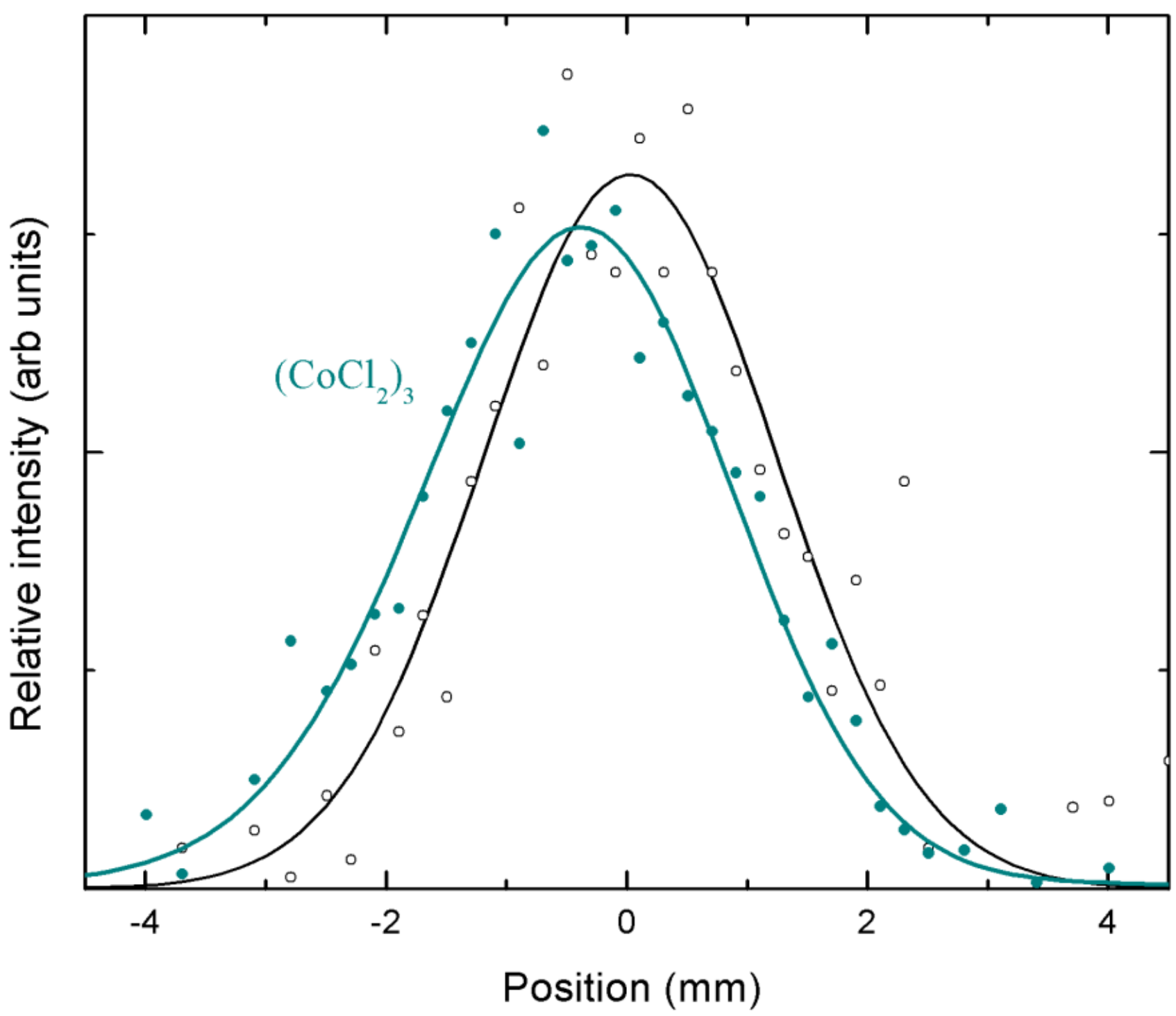

**Fig. S5.** Deflection profiles for nanodroplets doped with the trimer $(CoCl_2)_3$. The symbols are the same as in Fig. 1 in the main text. The profile was mapped out by setting the mass spectrometer detector to the $CoCl_3^+$ fragment ion peak. The average size of nanodroplets emitted by the nozzle in this experiment was $\approx 6\times10^3$ He atoms, while the average size of nanodroplets large enough to pick up and retain a trimer was approximately three times larger.

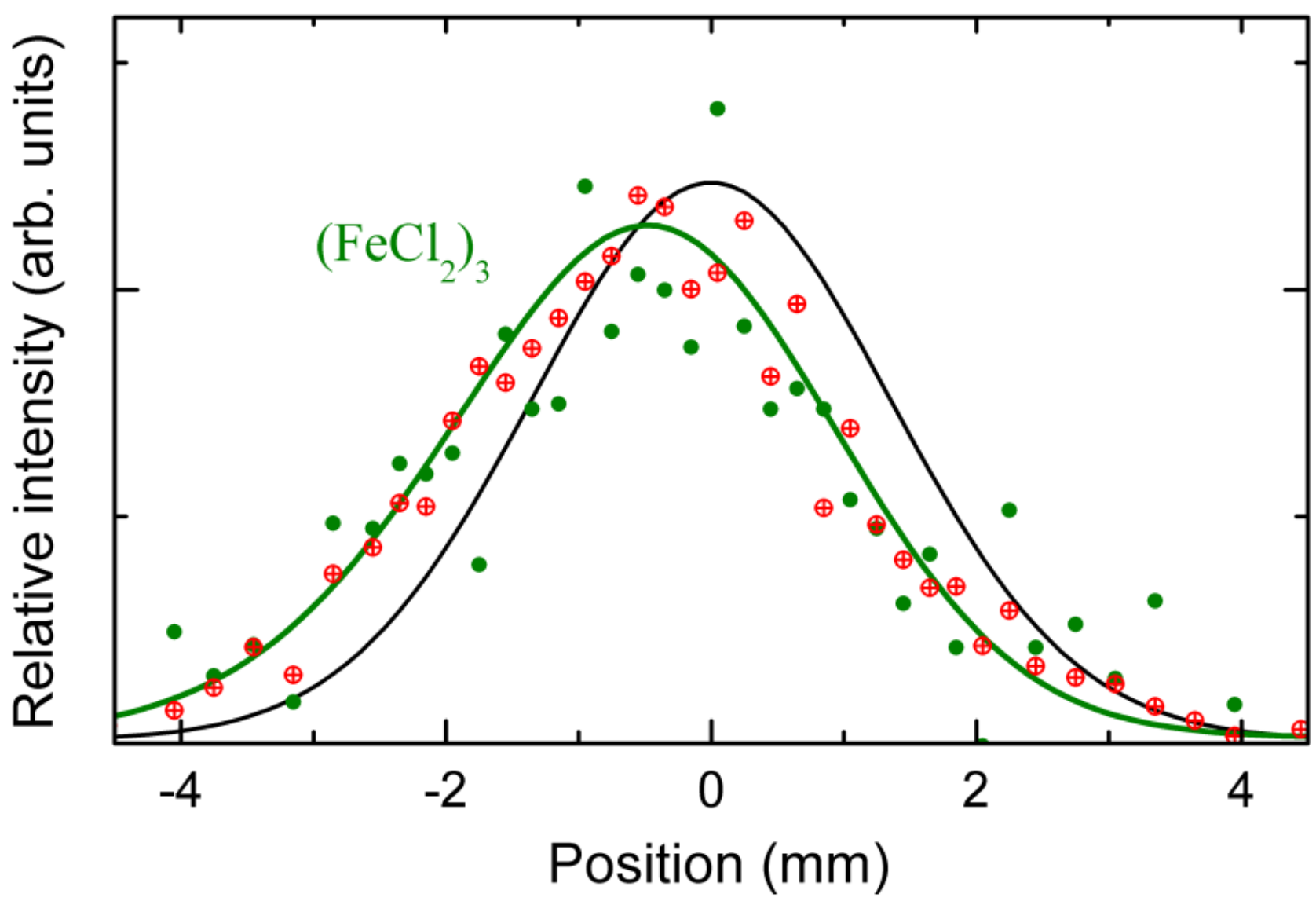

**Fig. S6.** Deflection of the $(FeCl_2)_3$ trimer measured by setting the mass spectrometer to two different ion fragments of the parent cluster. The solid dots represent the profile measured using the $FeCl_3^+$ peak, as in Fig. 2(B), while the crossed circles are derived from the $Fe_3Cl_5^+$ peak. The black centered line is the zero-field profile of the former fragment. Simulation of the smaller fragment deflection (green line) returned a value of $5.7\mu_B$, while for the larger fragment the result was $5.4\mu_B$. Within the $\pm 0.5\mu_B$ accuracy of the measurement, these values are in agreement.